\documentstyle[preprint,aps,eqsecnum,epsf,epsfig]{revtex}

\def\epsfig#1#2#3#4
         {
         \epsfysize=#2 \vbox{ \hglue#3 \epsfbox[#4]{#1} }
         }
\def\epsfigrot#1#2#3#4
         {
         \epsfxsize=#2
         \setbox\rotbox=\hbox to #2{\epsfbox[#4]{#1}}
         \vbox{\hglue#3 \rotl\rotbox}
         }
\newbox\rotbox
\begin{document}
\draft
\title{Finite Temperature Correlation Functions in Integrable QFT}
\author{A. LeClair$^a$, G. Mussardo$^{b,c}$}
\address{$^a$ Newman Laboratory, Cornell University, Ithaca, NY
14853.}
\address{$^b$ Dipartimento di Fisica, Universita' dell'Insubria, Como}  
\address{$^c$ Istituto Nazionale Fisica Nucleare, Sezione di Trieste}
\date{\today}
\maketitle

\begin{abstract}

Finite temperature correlation functions in integrable quantum field
theories are formulated only in terms of the usual, temperature-independent
form factors, and certain thermodynamic filling fractions
which are determined from the thermodynamic Bethe ansatz.  Explicit
expressions are given for the one and two-point functions.   
\end{abstract}
\pacs{~}
\vskip 0.2cm
\narrowtext
%
%
%
%

\newcommand{\appsection}{\addtocounter{section}{1} \setcounter{equation}{0}
                         \section*{Appendix \Alph{section}}}
\def\EQ{\begin{equation}}
\def\EN{\end{equation}}
\def\ch{{\rm ch}}
\def\sh{ {\rm sh}}
\def\oti{{\otimes}}
\def\bra#1{{\langle #1 |  }}
\def\lb{ \left[ }
\def\rb{ \right]  }
\def\tilde{\widetilde}
\def\bar{\overline}
\def\hat{\widehat}
\def\*{\star}
\def\[{\left[}
\def\]{\right]}
\def\({\left(}		\def\BL{\Bigr(}
\def\){\right)}		\def\BR{\Bigr)}
	\def\BBL{\lb}
	\def\BBR{\rb}
%
%
\def\zb{{\bar{z} }}
\def\zbar{{\bar{z} }}
\def\frac#1#2{{#1 \over #2}}
\def\inv#1{{1 \over #1}}
\def\half{{1 \over 2}}
\def\d{\partial}
\def\der#1{{\partial \over \partial #1}}
\def\dd#1#2{{\partial #1 \over \partial #2}}
\def\vev#1{\langle #1 \rangle}
\def\ket#1{ | #1 \rangle}
\def\rvac{\hbox{$\vert 0\rangle$}}
\def\lvac{\hbox{$\langle 0 \vert $}}
\def\2pi{\hbox{$2\pi i$}}
\def\e#1{{\rm e}^{^{\textstyle #1}}}
\def\grad#1{\,\nabla\!_{{#1}}\,}
\def\dsl{\raise.15ex\hbox{/}\kern-.57em\partial}
\def\Dsl{\,\raise.15ex\hbox{/}\mkern-.13.5mu D}
%
%
\def\th{\theta}		\def\Th{\Theta}
\def\ga{\gamma}		\def\Ga{\Gamma}
\def\be{\beta}
\def\al{\alpha}
\def\ep{\epsilon}
\def\vep{\varepsilon}
\def\la{\lambda}	\def\La{\Lambda}
\def\de{\delta}		\def\De{\Delta}
\def\om{\omega}		\def\Om{\Omega}
\def\sig{\sigma}	\def\Sig{\Sigma}
\def\vphi{\varphi}
%
%
\def\CA{{\cal A}}	\def\CB{{\cal B}}	\def\CC{{\cal C}}
\def\CD{{\cal D}}	\def\CE{{\cal E}}	\def\CF{{\cal F}}
\def\CG{{\cal G}}	\def\CH{{\cal H}}	\def\CI{{\cal J}}
\def\CJ{{\cal J}}	\def\CK{{\cal K}}	\def\CL{{\cal L}}
\def\CM{{\cal M}}	\def\CN{{\cal N}}	\def\CO{{\cal O}}
\def\CP{{\cal P}}	\def\CQ{{\cal Q}}	\def\CR{{\cal R}}
\def\CS{{\cal S}}	\def\CT{{\cal T}}	\def\CU{{\cal U}}
\def\CV{{\cal V}}	\def\CW{{\cal W}}	\def\CX{{\cal X}}
\def\CY{{\cal Y}}	\def\CZ{{\cal Z}}

\def\rvac{\hbox{$\vert 0\rangle$}}
\def\lvac{\hbox{$\langle 0 \vert $}}
\def\comm#1#2{ \BBL\ #1\ ,\ #2 \BBR }
\def\2pi{\hbox{$2\pi i$}}
\def\e#1{{\rm e}^{^{\textstyle #1}}}
\def\grad#1{\,\nabla\!_{{#1}}\,}
\def\dsl{\raise.15ex\hbox{/}\kern-.57em\partial}
\def\Dsl{\,\raise.15ex\hbox{/}\mkern-.13.5mu D}
%
%
%
\font\numbers=cmss12
\font\upright=cmu10 scaled\magstep1
\def\stroke{\vrule height8pt width0.4pt depth-0.1pt}
\def\topfleck{\vrule height8pt width0.5pt depth-5.9pt}
\def\botfleck{\vrule height2pt width0.5pt depth0.1pt}
\def\Zmath{\vcenter{\hbox{\numbers\rlap{\rlap{Z}\kern
0.8pt\topfleck}\kern 2.2pt
                   \rlap Z\kern 6pt\botfleck\kern 1pt}}}
\def\Qmath{\vcenter{\hbox{\upright\rlap{\rlap{Q}\kern
                   3.8pt\stroke}\phantom{Q}}}}
\def\Nmath{\vcenter{\hbox{\upright\rlap{I}\kern 1.7pt N}}}
\def\Cmath{\vcenter{\hbox{\upright\rlap{\rlap{C}\kern
                   3.8pt\stroke}\phantom{C}}}}
\def\Rmath{\vcenter{\hbox{\upright\rlap{I}\kern 1.7pt R}}}
\def\Z{\ifmmode\Zmath\else$\Zmath$\fi}
\def\Q{\ifmmode\Qmath\else$\Qmath$\fi}
\def\N{\ifmmode\Nmath\else$\Nmath$\fi}
\def\C{\ifmmode\Cmath\else$\Cmath$\fi}
\def\R{\ifmmode\Rmath\else$\Rmath$\fi}








\section{Introduction}

Integrable quantum field theories in $1+1$ space-time dimensions
are fundamentally characterized by their factorizable S-matrix 
\cite{Zam}. The S-matrix along with certain bootstrap axioms in
principle characterize the matrix elements of fields in the
space of asymptotic multi-particle states (Form Factors) 
\cite{Smirnov,Karowski}. Finally, the correlation 
functions at zero temperature can be characterized by 
infinite integral representations using the form factors.  
The S-matrix also governs the thermodynamics of the system 
through the Thermodynamic Bethe Ansatz (TBA) \cite{Alosh}.

For many applications, especially in Condensed Matter Physics 
(see, for instance \cite{Tsvelik}), the finite temperature 
correlation functions are of importance. In the Matsubara 
imaginary time formalism \cite{Matsubara}, finite temperature 
correlation functions can be viewed as correlation functions 
on an infinite cylindrical geometry, where the spacial 
coordinate $-\infty < x < \infty$ runs along the length of 
the cylinder and the euclidean time lives on a circle of 
radius $R= 1/T$, where $T$ is the temperature. In a picture 
where the hamiltonian evolution is along the circumference
of the cylinder, let us call it the $R$-channel, the Hilbert 
space lives on the infinite line of the coordinate $x$, thus 
the states of the Hilbert space one has to sum over in performing
thermodynamic averages have the usual multi-particle description 
in infinite volume.  Deceptively simple as it may appear, 
this is an important observation, since it implies that in 
this picture, the matrix elements of operators do not depend
on $R$, and can be expressed then by the usual, 
temperature-independent form factors. Formally, one has
\begin{equation}
\label{Rchan}
\langle \CO \rangle_R = \inv{Z} {\rm Tr} \( e^{-RH}  \CO \)  
~~~~~~~~({\rm R-channel}) 
\end{equation}
where $Z$ is the partition function. In this paper we use the 
$R$-channel to formulate the finite temperature correlation 
functions, the main goal being to use the Thermodynamic
Bethe Ansatz to make sense of (\ref{Rchan}). We show that 
the correlation functions can be characterized using only 
the usual form factors and some thermodynamic data, the so-called
filling fractions, which are  available
from the TBA. We present explicit integral representations for
the one and two point correlation functions.

Our formulation should be contrasted to the orthogonal picture, 
where hamiltonian evolution is viewed as along the length 
of the cylinder ($x$-direction).  In this ``$L$-channel'', 
the Hilbert space is on the finite volume $R$, and the 
computation of correlation functions does not involve the 
sum over thermodynamic states but rather involves the notion 
of the ground state $|0_R \rangle $ on the circle of
radius R 
\begin{equation} 
\label{Lchan} 
\langle \CO \rangle_R = \langle 0_R | \CO | 0_R \rangle  
~~~~~~~({\rm L-channel})
\end{equation}
Presently, the precise structure of this ground state has not  
been identified  and moreover, in this $L$-channel, matrix 
elements of operators may depend on $R$ in a complicated unknown way.
There has been however some progress in this direction by 
Smirnov.  In fact, in a series of papers \cite{Smirnov2}, he 
studied properties of such matrix elements in a semi-classical limit. 

\section{One-Point Functions} 

Since the main features of the finite temperature calculation of 
correlation functions are already present in the simplest case 
of the one--point function of a quantum field operator $\CO$, let us start 
the discussion from the analysis of this case. Throughout this paper 
we let $R=1/T$, where $T$ is the temperature. To simplify the following 
discussion, we assume that in the field theory in question there is one
kind of particle $A$ of mass $m$. Generalization to other cases is quite easy.
Multi-particle states are denoted as $|\theta_1, \cdots, \theta_n \rangle$, 
where $\theta$ is the rapidity parameterizing the energy and momentum
\begin{equation}
\label{2.1}
e(\theta) = m \, \ch \theta , ~~~~~ k(\theta) = m \, \sh \theta \,\,\,.
\end{equation}
Using the resolution of the identity, 
\begin{equation}
\label{2.2}
1 = \sum_{n=0}^\infty \inv{n!} \int 
\frac{d\theta_1}{2\pi} \cdots \frac{d\theta_n}{2\pi} 
~ |\theta_1 , \cdots , \th_n \rangle \langle \th_n , \cdots , \th_1 | 
\,\,\, ,
\end{equation}
the one-point function of a local field $\CO(x,t)$ at finite temperature
has the formal representation
\begin{equation}
\label{2.3}
\langle \CO (x,t) \rangle_R = \inv{Z} 
\sum_{n=0}^\infty \inv{n!} \int 
\frac{d\theta_1}{2\pi} \cdots \frac{d\theta_n}{2\pi} 
\( \prod_{i=1}^n  e^{-e(\theta_i ) R } \) 
\langle \th_n , \cdots, \th_1 | \CO (0,0) | \th_1 , \cdots, \th_n \rangle 
\,\,\, ,
\end{equation}
where $Z = {\rm Tr} \left(e^{-RH}\right)$ is the partition function. From
translation  invariance, the above quantity does not depend on $x,t$ and 
 is only a function of the scaling variable $m R$.  

Our aim is to show that the previous expression can  
actually be  organized in a way which better reveals its physical content 
and which moreover provides an efficient method for its actual 
computation. In addition, we will also show that the final form of the 
one--point functions may be regarded as a generalization of the
formula obtained in the case of a free theory, once an appropriate
dictionary between the interacting and free theories has been 
established. For this reason, let us initially consider the case of a
free fermionic  model\footnote{In this paper we consider 
only ``fermionic'' theories with $S(\theta = 0) = -1$. The formal
extension to bosonic theories with $S(\theta = 0) = 1$ is a simple 
exercise.}, i.e. an integrable model with S-matrix $S=-1$. This case 
was originally studied in \cite{LLSS}. For the following
considerations, we need in particular to recall the main 
findings reported in the appendix A of \cite{LLSS}. 

\subsection{Main results in the free case} 

In carrying out the sum over states it is important to include
$\delta$-function contributions to the form factors as follows.  
Let $\CA = \{ \th_n \cdots \th_1 \} $ and $\CB = \{ \th'_1 \cdots 
\th'_m \} $ denote some ordered sets of rapidities.   Then the
form factor is expressed as a sum over all ways of breaking up 
$\CA$ and $\CB$ into two sets: 
\begin{equation}
\label{2.4} 
\langle \CA | \CO(x,t) | \CB \rangle = 
\sum_{\CA = \CA_1 \cup \CA_2 ; \CB = \CB_1 \cup \CB_2 } 
S_{\CA, \CA_1} S_{\CB , \CB_1} \langle \CA_2 | \CB_2 \rangle 
\langle \CA_1 | \CO(x,t) | \CB_1 \rangle_{\rm conn} \,\,\, , 
\end{equation} 
where
$S_{\CA, \CA_1}$ are S-matrix factors required to bring $|\CA \rangle $
into the order $| \CA_2 , \CA_1 \rangle$, i.e. 
$\langle \CA | = S_{\CA , \CA_1 } \langle \CA_2 , \CA_1 | $, and
similarly for $|\CB \rangle$. 
The inner products $\langle \CA_2 | \CB_2 \rangle$ are most easily
evaluated by introducing free particle creation-annihilation
operators $| \th_1 \cdots \th_n \rangle = A^\dagger (\th_1 ) \cdots
A^\dagger (\th_n ) | 0 \rangle$, and using 
$\{ A(\th) , A^\dagger (\th' ) \} = 2\pi \delta (\th - \th' )$;  
they are sums of products of $\delta$-functions. 
The ``connected'' piece of the form factor 
$\langle \CA_1 | \CO (x,t) | \CB_1 \rangle_{\rm conn} $ is defined
to be the form factor with no overlap, i.e. $\langle \CA_1 | \CB_1 \rangle 
= 0$.  The crossing relation is thus valid and one can define 
\begin{equation}
\label{2.5} 
\langle \th_n \cdots \th_1 | \CO  | \th'_1 \cdots \th'_m 
\rangle_{\rm conn} 
\equiv {\large FP}
\( \lim_{\eta_i \to 0 } 
\langle 0 | \CO  | \th'_1 \cdots \th'_m , 
\th_n -i\pi + i \eta_n , \cdots , \th_1 -i\pi + i \eta_ 1 \rangle \) \,\,\,
\end{equation} 
where $FP$ in front of the expression means taking its {\em finite part}, 
i.e. terms proportional to $(1/\eta_i)^p$, where $p$ is some positive power,
and also terms proportional to 
$\eta_i / \eta_j $, $i\neq j$ are discarded
in taking the limit. With this prescription the resulting
expression  is
independent of the way in which 
the above limits are taken, and it is therefore the only quantity 
which has an unambiguous physical meaning. This way of taking 
the limit has been already used in the literature (see \cite{Balog}). 

\def\vep{\varepsilon}

There are two effects of the $\delta$-function terms coming from 
$\langle \CA_2 | \CB_2 \rangle$ when $\CA \neq \CA_1$. The first 
effect is due to terms involving $[\delta (\theta - \theta ' )]^2 $;
these give rise to an  overall factor which sums up to the partition 
function $Z$, and therefore cancels out from the final result. The 
other effect of the $\delta$-functions is to give rise to a
``filling-fraction'' $f(\theta)$ for integrations over rapidity. 
The final expression of the one--point functions is then 
given by 
\begin{equation} 
\label{2.6}
\langle \CO (x,t) \rangle_R  
= \sum_{n=0}^\infty \inv{n!} 
\int \frac{d\th_1}{2\pi} \cdots \frac{d\th_n}{2\pi} 
\( \prod_{i=1}^n  f(\th_i ) e^{-\vep (\th_i) } \) 
\langle \th_n \cdots \th_1 | \CO (0) | \th_1 \cdots \th_n \rangle_{\rm conn} 
\, ,
\end{equation} 
where here $\vep = e(\th) R$, the connected form factor is defined 
in (\ref{2.5}),  and 
\begin{equation}
\label{2.7}
f(\theta ) = \inv{ 1 + e^{-\vep (\th)} } \,\,\, .
\end{equation} 
In the following it will be  convenient to introduce the functions 
\EQ
\label{fillsig}
f_\sigma (\th) =  \inv{1+ e^{-\sigma \vep(\th)}} \,\,\,.
\EN
so that $f = f_+ $ and $ f e^{-\vep} = f_-$. 
As we will show below, the expression (\ref{2.6}) for the one--point 
function  is the one which generalizes to the interacting case. 

\def\phi{\varphi}

\subsection{Interacting case. Quasi--particle excitations}

Let us now turn to the interacting case.  Again, for simplicity, we 
assume the theory has a spectrum given by a single particle $A $ 
of mass $m$ with an S-matrix $S(\theta)$.  
We define 
\begin{equation}
\label{2.8} 
\sigma (\theta) = -i \log S (\theta) , ~~~~~
\phi (\theta) = -i \frac{d}{d\theta} \log S (\theta) \,\,\,. 
\end{equation} 
The partition function at a finite temperature $T$ and on a volume $L$ 
(for $L \rightarrow \infty$) is determined by means of the
the Thermodynamic Bethe Ansatz equations as follows 
\cite{Alosh,Yang}. In a box of large volume $L$, $0<x<L$, 
the quantization condition of the momenta is given by $e^{ik(\th_i ) L }
\prod_{j\neq i}  S(\th_i - \th_j ) = 1$, which can be equivalently 
expressed as 
\begin{equation}
\label{2.9}  
m L \,\sh \th_i + \sum_{j\neq i} \sigma (\th_i - \th_j ) = 2\pi n_i
\,\,\, ,  
\end{equation}
where $n_i$ are integers.  Introducing a density of occupied states
per unit volume $\rho_1 (\theta)$ as well as a density of levels 
$\rho (\theta) $, in the thermodynamic limit eq.\,(\ref{2.9}) becomes 
\begin{equation} 
\label{2.10}
2\pi \rho = e + 2\pi \phi * \rho_1 \,\,\, , 
\end{equation}
where $(f*g)(\th ) = \int_{-\infty}^\infty d\th' f(\th - \th') 
g(\th') /2\pi $.  Defining the pseudo-energy $\vep (\th)$ as 
\begin{equation}
\label{2.11} 
\frac{\rho_1}{\rho} = \inv{1+ e^{\vep} } \,\,\, ,
\end{equation}
the minimization of the free-energy with respect to the densities of 
states leads to the integral equation  
\begin{equation}
\label{2.12} 
\vep = e R - \phi * \log ( 1 + e^{-\vep}  ) \,\,\, ,
\end{equation}
and the partition function is then given by 
\begin{equation}
\label{2.13} 
Z (L, R) = \exp\left[ mL 
\int \frac{d\th}{2\pi} \ch \th ~ \log 
\( 1 + e^{-\vep (\th )} \)\right] \,\,\, .
\end{equation} 

The interesting point is  that the above partition function can be
interpreted as  one of a {\it free} gas of fermionic particles 
but with energy given by $\vep(\th)/R$. Namely, there is a one--to--one 
correspondence between the above expression (\ref{2.13}) and a  
partition function computed according to the following sum 
\begin{equation}
\label{2.14}
Z(L,R) = \sum_{n=0}^\infty \inv{n!} 
\int \frac{d\th_1}{2\pi} \cdots \frac{d\th_n}{2\pi} 
~ \langle \th_n \cdots \th_1 | \th_1 \cdots \th_n \rangle ~  
\prod_{i=1}^n e^{-\vep (\th_i ) } \,\,\, ,
\end{equation}
where the scalar products of the states are computed by applying the 
standard {\em free} fermionic rules. To see this, let us define 
\begin{equation}
F(R) = 
\int \frac{d\th}{2\pi} \ch \th ~ \log \( 1 + e^{-\vep (\th )} \) \,\,\, ,
\label{freeenergy}
\end{equation} 
which admits the series expansion 
\begin{equation}
F(R) = \sum_{n=0}^{\infty} \frac{(-1)^{n+1}}{n} \,I_n(R) \,\,\, , 
\label{expansionfree}
\end{equation}
where 
\begin{equation}
I_n(R) \equiv  \int \frac{d\th}{2\pi} \ch \th \,e^{-n \vep (\th)} \,\,\,.
\end{equation}
The partition function obtained by the TBA equation has the 
following expansion in powers of $(m L)$ 
\begin{equation}
Z(L,R) = 1 + (m L) F(R) + \frac{(m L)^2}{2!} (F(R))^2 + \cdots 
\frac{(m L)^n}{n!} (F(R))^n + \cdots 
\label{series1}
\end{equation}

Let us now compute the partition function by using the other expression 
(\ref{2.14}), with a regularization of the squares of the $\delta$-functions
appearing in $\langle \th_n \cdots \th_1 | \th_1 \cdots \th_n \rangle$
as for a free fermionic theory on a sufficiently large volume $L$: 
\begin{equation}
\label{2.15}
\[ \delta (\th - \th' ) \]^2 \equiv \frac{mL}{2\pi} \,\ch (\th ) \, 
\delta (\th - \th' ) \,\,\,.
\end{equation} 
Hence 
\begin{equation}
Z(L,R) = 1 + Z_1 + Z_2 + \cdots Z_n + \cdots 
\end{equation}
where for the first terms we have 
\begin{eqnarray}
Z_1 &=& \int \frac{d\th}{2\pi} <\th | \th> e^{-\vep(\th)} 
= \int \frac{d\th}{2\pi} d\th' \delta(\th - \th' ) <\th' | \th> 
e^{-\vep (\th) } = 
\label{2.16} 
\\
\,\,\,&\,\,\, = & mL \int \frac{d\th}{2\pi} \ch (\th) e^{-\vep (\th )} = (m L)
\, I_1  \,\,\, ;
\nonumber
\end{eqnarray}
\begin{eqnarray}
Z_2 &=& 
\frac{1}{2} \int \frac{d\th_1}{2\pi} \frac{d\th_2}{2\pi} 
<\th_2 \th_1 | \th_1 \th_2 > e^{-\vep(\th_1) - \vep(\th_2)} = 
\frac{1}{2} \int \frac{d\th_1}{2\pi} \frac{d\th_2}{2\pi} 
\left[ (2\pi)^2 ( 
\delta(\th_1-\th_1) \delta(\th_2-\th_2) + \right. \nonumber\\
&& \,\,\,\left. - \delta(\th_1-\th_2) 
\delta(\th_2-\th_1))\right] e^{-\vep(\th_1) - \vep(\th_2)} = 
\frac{1}{2} (m L)^2 I_1^2 - \frac{1}{2} (m L) I_2 \,\,\, ,
\label{Z2}
\end{eqnarray}
and for the next few   
\begin{equation}
Z_3 = \frac{(m L)^3}{3!} I_1^4 - \frac{(m L)^2}{2} I_1 I_2 + 
\frac{(m L)}{3} I_3 \,\,\, ,
\end{equation}
\begin{equation}
Z_4 = \frac{(m L)^4}{4!} I_1^4 - (m L)^3 I_1^2 I_2 + 
\frac{(M L)^2}{2} \left[\frac{2}{3} I_1 I_3 + \left(\frac{I_2}{2}\right)^2
\right] - \frac{(m L)}{4} I_4 \,\,\, . 
\end{equation}
It is not difficult to carry on to higher order  
and  to show that the series (\ref{2.14}) precisely coincides 
with the one of eq.\,(\ref{series1}), the only difference being
in the arrangement of the single terms. For instance, collecting
the terms proportional to $(m L)$ in all $Z_n$, one simply obtains for 
their sum $F(R)$, whereas the sum of the terms proportional 
to $(m L)^n$ appearing in all $Z_n$ gives rise to the higher 
power $(F(R))^n$.  

The above remarkable fact can be interpreted as meaning that 
all physical properties of the system can be extracted by the 
quasi-particle excitations above the TBA thermal ground state. 
These excitations have dressed energy $\tilde{e} = \vep(\th) /R$ and 
dressed momenta $\tilde{k} (\th)$:  
\begin{equation}
\label{2.17} 
\tilde{e} (\th ) = \vep (\th )/R \, , ~~~~~~
\tilde{k} (\th ) = k(\th) + 2\pi (\sigma * \rho_1 ) (\th ) \,\,\,.  
\end{equation}
In this context, the rapidity $\th$ plays then the role 
of a variable which simply parameterizes the dispersion relation 
of the quasi--particle excitations. This result was already understood 
in a non--relativistic situation by Yang and Yang \cite{Yang} and 
a generalization of their result in our relativistic context is 
given for convenience in appendix A.  

In the light of these considerations, let us return now to 
the sum (\ref{2.3}). It is now clear that if we replace 
$e(\th)$ by $\tilde{e}(\th )$ and proceed to evaluate the 
sum as in the free theory using (\ref{2.4}), the partition 
function will properly be factored out and be canceled by 
the $Z$ in the denominator, and furthermore the filling
fractions will again be generated with $\vep$ now given by the 
TBA pseudo-energy. Thus, one obtains precisely the same 
formulae as (\ref{2.6}, \ref{2.7}). We emphasize that the 
connected form factor appearing in (\ref{2.6}) is the 
usual (temperature-independent) form factor. Then the 
interpretation of eq.\,(\ref{2.6}) is quite clear, i.e. 
each multiparticle state contributes to the thermal sum 
with its $T=0$ form factor but each is  weighted with 
the thermal probability of the occupation of that state, 
as determined self--consistently by the TBA equations.  

A non-trivial check of the above claim is readily available for
the trace of the energy momentum tensor $\CO = T_\mu^\mu$. The connected 
form-factors of this operator can be easily extracted and they depend 
only on the model through the function $\phi$. For the first ones we have 
\begin{eqnarray}
\label{2.18} 
\langle \th | T_\mu^\mu | \th \rangle_{\rm conn} &=& 2\pi m^2 \,\,\, ; 
\\ \nonumber
\langle \th_2 , \th_1 | T_\mu^\mu | \th_1 , \th_2 \rangle_{\rm conn} 
&=& 4\pi m^2 \phi(\th_1 - \th_2 ) \ch (\th_1 - \th_2 ) \,\,\, ,
\end{eqnarray}
and inductive application of the form factor residue equations 
leads to 
\begin{eqnarray}
\label{2.19} 
\langle \th_n \cdots \th_1 | T_\mu^\mu | \th_1 \cdots \th_n \rangle_{\rm
conn} & = & 2\pi m^2 \, \phi (\th_{12} ) \phi (\th_{23}) \,\,\, ,
\cdots \phi(\th_{n-1,n})\,\ch (\th_{1n} )  \\
& & ~~~~~~~~~~~~+{\rm  permutations}  
\end{eqnarray}
where $\th_{ij} = \th_i - \th_j $. For the aim of computing the 
thermal trace (i.e. considered as inserted into integral expressions  
over $\th_i$), the last expression can be taken as  
\EQ
\langle \th_n \cdots \th_1 | T_\mu^\mu | \th_1 \cdots \th_n \rangle_{\rm
conn}  =  2\pi m^2 \,n!\, \phi (\th_{12} ) \phi (\th_{23}) \,\,\, ,
\cdots \phi(\th_{n-1,n})\,\ch (\th_{1n} ) \,\,\,.
\EN 
Finally, the vacuum expectation value 
of this operator is given by 
\begin{equation}
<0 | T_{\mu}^\mu | 0> \equiv (T_\mu^\mu)_0 
= \frac{\pi m^2}{2 \sum_i \sin\pi\alpha_i} \,\,\, , 
\end{equation}
where $\alpha_i$ are the resonance angles entering the two--body 
scattering amplitude of the particle (in the more general case, 
one should consider the scattering amplitude of the particle with 
the lightest mass). Therefore 
we have 
\begin{equation} 
\label{2.25} 
\langle T^\mu_\mu \rangle_R = (T_\mu^\mu)_0 + 2\pi m^2 
\( \sum_{n=1}^\infty \int 
\[ \prod_{i=1}^n   \frac{d\th_i}{2\pi} f_- (\th_i ) \]  
\phi (\th_{12} ) \cdots \phi(\th_{n-1, n} )\,\ch (\th_{1n} ) 
\) \,\,\, . 
\label{seriestrace}   
\end{equation}
The $n$-th term of this series can be represented by the  
graph 
\vspace{1cm}
\begin{center}
\begin{minipage}[b]{.65\linewidth}
\centerline{\epsfxsize=6.cm\epsfbox{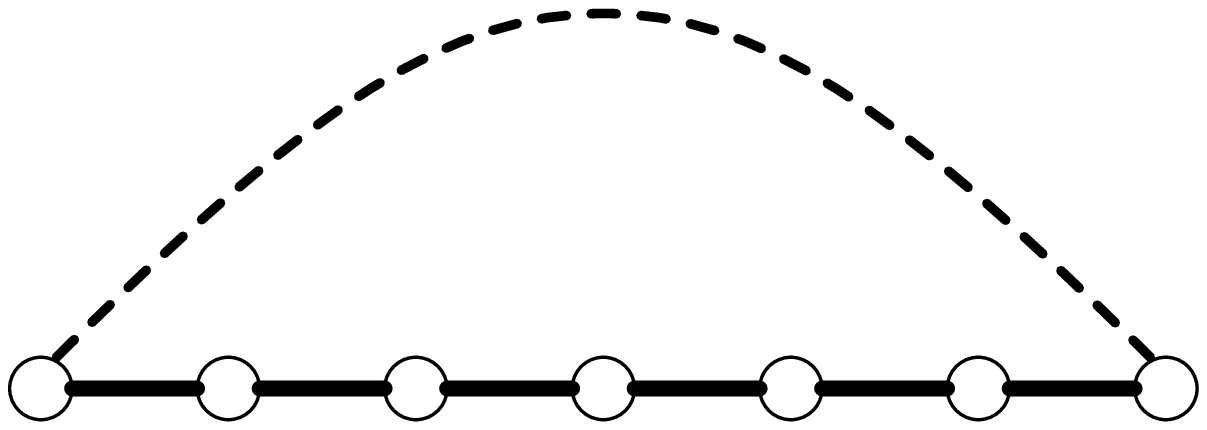}} 
\end{minipage}
\end{center}
where the dots are associated to the weights $f_-(\th_i)$, 
the thick lines between the dots to the functions 
$\phi(\th_i-\th_{i+1})$ and the dotted line which links 
the ending dots to $\ch(\th_{1n})$.  

On the other hand, $\langle T_\mu^\mu \rangle_R$ can be computed
directly from the TBA \cite{Alosh} as 
\EQ
\langle T_\mu^\mu \rangle_R - (T_\mu^\mu)_0
= \frac{2\pi}{R}  \frac{d}{dR}  [RE(R)] \,\,\, ,
\EN
where $E(R) = -\log Z /L$.  This can be  expressed as 
\begin{equation}
\label{2.20} 
\langle T_\mu^\mu \rangle_R - (T_\mu^\mu)_0 = m \int d\th 
\frac{e^{-\vep}}{1+ e^{-\vep} } 
\(  \d_R \vep \,  \ch \th  - \inv{R}  \d_\th \vep \, \sh \th \)
\,\,\, , 
\end{equation}
where the functions $\d_R \vep$ and $\d_\th \vep$ satisfy linear integral
equations which can be easily solved. In fact, define the kernel 
\begin{equation} 
\label{2.21} 
\hat{K} (\th , \th') = \phi(\th, \th') 
\frac{e^{-\vep(\th')}}{1+ e^{-\vep (\th' )} } \,\,\, .  
\end{equation}
Then, by differentiating the integral equation (\ref{2.12}) one obtains
\begin{equation}
\label{2.22}
(1-\hat{K}) \d_R \vep = e , ~~~~~ (1-\hat{K}) \inv{R} \d_\th \vep = k 
\,\,\, , 
\end{equation}
where 
$(\hat{K} \d_R \vep ) (\th ) \equiv \int d\th' 
\hat{K} (\th , \th' ) \d_R \vep (\th' ) / 2\pi $.  
Introducing the resolvent $\hat{L}$ satisfying 
\begin{equation}
\label{2.23b} 
(1+\hat{L}) (1- \hat{K} ) = 1 \,\,\, ,
\end{equation} 
we have 
\begin{equation}
\label{2.23c} 
\d_R \vep = (1+\hat{L}) e ~~~, ~~~~~
\inv{R} \d_\th \vep = (1+ \hat{L}) k \,\,\, . 
\end{equation} 
Using $1+\hat{L} = \sum_{n=0}^\infty \hat{K}^n $, one can easily express
the above in an integral series:
\begin{equation}
\label{2.24}
\d_R \vep = e + \phi* \( \frac{e^{-\vep}}{1+e^{-\vep}} e \) + ... 
\end{equation}
and similarly for $\d_\th \vep$.  Substituting these series into 
(\ref{2.20}) one then finds a perfect agreement with expression
(\ref{2.25}). 

A model with  one particle whose computation of one--point functions
can be explicitly performed is provided by the Sinh-Gordon model. This 
is briefly discussed in the next section. 

\subsection{One--point functions in the Sinh--Gordon model} 

The Sinh-Gordon theory is defined by the action 
\EQ
{\cal S}\,=\,\int d^2x \left[
\frac{1}{2}(\partial_{\mu}\Phi)^2-\frac{m_0^2}{g^2}
\,\cosh\,g\Phi(x)\,\right]\,\,.
\label{Lagrangian}
\EN
The model is invariant under a $Z_2$ symmetry $\Phi\rightarrow -\Phi$. 
Its two-particle $S$-matrix is given by 
\EQ
S(\th,\alpha)\,=\,
\frac{\tanh\frac{1}{2}(\th - i\pi \alpha)}
{\tanh\frac{1}{2}(\th + i\pi \alpha)} \label{smatrix}\,\, ,
\EN
where $\alpha$ is the so-called renormalized coupling constant
$
\alpha(g)\,=\,\frac{g^2}{8\pi} \frac{1}{1+\frac{g^2}{8\pi}}.
$
For real values of $g$ the $S$-matrix has no poles in the physical sheet 
and hence there are no bound states. The function $\varphi(\th)$ of 
this model is explicitly given by 
\EQ
\varphi(\th) = \frac{2 \sin\alpha \cosh\th}{\cosh^2\th - \cos^2\pi\alpha}
\,\,\,.
\EN 
We are interested, in particular, in computing the one--point 
functions of the exponential operators of this theory, 
$\Psi_k(x)=e^{k g \Phi}$. To this aim, let us recall the general 
structure of the form-factors of the Sinh-Gordon model, as determined in 
\cite{FMS,KM}, specializing at the end to the operators in question. 

For scalar operators, the form--factors of the theory can be
parameterized as follows
\EQ
F_n(\th_1,\ldots,\th_n)\,=\, H_n\, Q_n(x_1,\ldots,x_n)\,
\prod_{i<j} \frac{F_{\rm min}(\th_{ij})}{(x_i+x_j)}
\sb ,\label{para} \,\,\,,
\EN
where $x_i\equiv e^{\th_i}$ and $\th_{ij}=\th_i-\th_j$. $F_{\rm min}(\th)$
is an analytic function explicitly given by 
\EQ
F_{\rm min}(\th,\alpha)\,=\,
\prod_{k=0}^{\infty}
\left|
\frac{\Gamma\left(k+\frac{3}{2}+\frac{i\hat\th}{2\pi}\right)
\Gamma\left(k+\frac{1}{2}+\frac{\alpha}{2}+\frac{i\hat\th}{2\pi}\right)
\Gamma\left(k+1-\frac{\alpha}{2}+\frac{i\hat\th}{2\pi}\right)}
{\Gamma\left(k+\frac{1}{2}+\frac{i\hat\th}{2\pi}\right)
\Gamma\left(k+\frac{3}{2}-\frac{\alpha}{2}+\frac{i\hat\th}{2\pi}\right)
\Gamma\left(k+1+\frac{\alpha}{2}+\frac{i\hat\th}{2\pi}\right)}
\right|^2 \,\,\, , 
\EN
where $\hat\th = i\pi -\th$. It satisfies the functional equations 
\begin{eqnarray} 
&& F_{\rm min}(\th) =F_{\rm min}(-\th)\, S(\th,\alpha)\,\, , 
\nonumber \\
&& F_{\rm min}(i\pi-\th) =F_{\rm min}(i\pi+\th)\,\, , 
\label{functionalequaF}\\
&& F_{\rm min}(i\pi+\th,\alpha) F_{\rm min}(\th,\alpha) =
\frac{\sinh\th}{\sinh\th+\sinh(i\pi \alpha)}\,\,\, .
\nonumber 
\end{eqnarray}
where the last equation is particularly important for the computation 
at a finite temperature. $H_n$ are normalization constants, which can 
be conveniently chosen as $H_{2n+1} = H_{1} \mu^{2n}$, $H_{2n} = H_{2}
\mu^{2n-2}$,  with 
$ \mu \equiv 
\left ( \frac{4 \sin (\pi \alpha)}{F_{min}(i\pi,\alpha)}
\right )^{\frac 12} 
$.  
In absence of additional requirements on the form--factors, 
$H_1$, $H_2$ are two independent parameters. Finally, the functions
$Q_n(x_1,\dots,x_n)$ are symmetric polynomials in the variables $x_i$, 
which are fixed by the residue equations satisfied by the form factors 
and whose total and partial degrees depend on the specific operator.  
In general, they can be expressed in terms of the elementary symmetric
polynomials $\sigma^{(n)}_k(x_1,\dots,x_n)$, defined by the generating
function 
\EQ
\prod_{i=1}^n(x+x_i)\,=\,
\sum_{k=0}^n x^{n-k} \,\sigma_k^{(n)}(x_1,x_2,\ldots,x_n)\sb . 
\label{generating}
\EN
Conventionally the $\sigma_k^{(n)}$ with $k>n$ and with $n<0$ are zero.
It is also convenient to introduce the symbol $[n]$ defined by 
$ 
[n] \equiv \frac{\sin ( n \pi\alpha)}{\sin \pi\alpha} 
$. 
As shown in \cite{KM}, an interesting class of solutions of the 
recursive equations satisfied by the Form Factors (FF)
is given by the following polynomials $Q_n$,  
\EQ
Q_{n}(k) = ||M_{ij}(k) || \,\, ,
\EN
where $M_{ij}(k)$ is an $(n-1)\times (n-1)$ matrix with 
entries\footnote{For simplicity the dependence of $Q_n(k)$ on the
variables $x_i$ has been suppressed.} 
\EQ
M_{ij}(k) =
\sigma_{2i-j}\, [i-j+k] \,\,\, , 
\label{element}
\EN
and the vertical lines denote the determinant of that matrix. 
These polynomials depend on an arbitrary number $k$ and satisfy 
$
Q_{n}(k)\,=\, (-1)^{n+1} Q_{n}(-k) \label{pr1}
$. 
Form--factors of several operators can be expressed in terms of the 
polynomials $Q_n(k)$. For instance, the whole set of FF of the 
elementary field $\Phi(x)$ are given by $Q_n(0)$, with
the choice $H_1=1/\sqrt{2}$, $H_2=0$ whereas those of $T_\mu^\mu(x)$,
are given by the even polynomials $Q_{2n}(1)$, with the choice 
$H_1=0$, $H_2 = 2\pi m^2$. 

The form--factors of the exponential operators $\Psi_k(x)=e^{k g \Phi}$, 
are just given by the $Q_n(k)$, with the normalization constants 
fixed to be $H_1^k = \mu [k]$, $H_2^k = \mu^2 [k]$. 
Since the one--point function of operators which are odd under the 
$Z_2$ symmetry $\Phi \rightarrow -\Phi$ is identically zero, 
the one--point correlators at finite temperature involving 
the exponential fields coincide  with those of the even combination
$\Psi_k + \Psi_{-k}$. Using the prescription (\ref{2.5}) one finds
\begin{eqnarray}
&& \frac{\langle \cosh \,k g \Phi \rangle_R}
{\langle 0|\cosh \,k g \Phi|0\rangle} 
= 
1 + 4 [k]^2 \sin\pi\alpha \int \frac{d\th}{2\pi} f_-(\th)
+ 4 [k]^2 \sin\pi\alpha \times 
\nonumber \\
&& \,\,\,\,\,\times  
\int \frac{d\th_1}{2\pi}
\frac{d\th_2}{2\pi} f_-(\th_1) f_-(\th_2)
\varphi(\th_{12}) \left( [k]^2  \cosh\th_{12} 
-\frac{[k-1][k+1]}{\cosh\th_{12}} \right) + \\
&& \,\,\,\, 
+ \int  
\frac{d\th_1}{2\pi}
\frac{d\th_2}{2\pi}
\frac{d\th_3}{2\pi} f_-(\th_1) f_- (\th_2) f_- (\th_3 ) 
\left( 4 \sin\pi\alpha [k]^6 \,\varphi(\th_{12}) \varphi(\th_{23})
\cosh\th_{13}  + \right. \nonumber \\
&& \,\,\,\, + \left. 
{\cal W}(\th_{12},\th_{13},\th_{23},k)\right) + \cdots 
\nonumber 
\end{eqnarray}
where the function ${\cal W}(\th_{12},\th_{13},\th_{23})$ appearing 
in the $3$--particle contribution can be expressed as  
\begin{eqnarray}
&& {\cal W}(\th_{12},\th_{13},\th_{23},k)  =   
\frac{4}{3} \, \varphi(\th_{12}) \varphi(\th_{13}) 
\varphi(\th_{23}) \times \nonumber 
\\
&& \,\,\,\,\, 
\left[ A(k) + B(k) \,\,\frac{1}{\cosh\th_{12} \cosh\th_{13}
\cosh\th_{23}}\,\,\,  + \right.
\\ && \,\,\,\,\,  \left.
+ \,\,\, C(k) \,\,
\frac{\cosh(\th_{13}-\th_{32}) + \cosh(\th_{12}-\th_{23}) + 
\cosh(\th_{21}-\th_{13})}
{\cosh\th_{12} \cosh\th_{13} \cosh\th_{23}} \right] \,\,\, ,\nonumber 
\end{eqnarray}
with the constants $A(k)$, $B(k)$ and $C(k)$ given by 
\begin{eqnarray*}
 A(k) & =&  \frac{7 [k]^2}{2} \left(
[k-1]^2 [k+1]^2 - [k-1] [k]^2 [k+1] \right) + 
\frac{[k]^3}{2} \left( [k-2] [k+1]^2 + [k-1]^2 [k+2] +
[k]^3 [2]^2 \right) \\
&&  -\frac{[k]}{8} \left(
[k-2] [k-1] [k+1]^3 + [k-2] [k]^3 [k+2] + 
[k-1]^3 [k+1] [k+2] - [k]^5 (1 - [2]^2)\right) 
\\
B(k)  &=&  \frac{[k]}{16} \left(
[k-2] [k]^2 [k+1]^2 + [k-1]^2 [k]^2 [k+2] + 2 
[k-1]^2 [k] [k+1]^2 + [k]^5 + \right.\\
&& - [k-2] [k-1] [k+1]^3 - [k-2] [k]^3 [k+2] - 
[k-1]^3 [k+1] [k+2] + \\
&& \left. - [k-1] [k]^3 [k+1] -3 [k-2] [k-1] [k] [k+1] [k+2] \right) 
\end{eqnarray*}
$$
\hspace{-11cm}
 C(k)   =  -\frac{1}{8} [k-1] [k]^4 [k+1] 
$$
The expressions for higher terms are rather complicated and are not 
given here. However, one can easily check that for $k = \pm 1$, the 
above expression coincides with the corresponding series 
of the trace of the stress--energy tensor\footnote{For the
check of the $3$--particle contribution one needs the trigonometric 
identity $[3]=-1+ [2]^2$.}. The plots of some of these quantities 
for some values of the coupling constant are shown in Figure 1. 
Notice that if the operator $\Psi_k$ is ``on resonance'' with the 
value of the coupling constant, namely if $k \alpha =n$, where 
$n$ is an integer, its one--point function coincides with its 
vacuum matrix element only. Moreover, varying $\alpha$, there may 
be a swapping of the profiles associated to the one--point
functions of different $\Psi_k$. 

\subsection{Low and high--temperature limits of the one--point functions}

Let us discuss in more detail the behavior of the one--point 
correlators (\ref{2.6}) as functions of the temperature. 
First notice that if the anomalous dimension of the operator $\CO$ 
is given by $2 \Delta_{\CO}$, its vacuum expectation value may be 
expressed as $\langle 0|\CO|0\rangle \equiv \langle \CO\rangle_0= Y
m^{2\Delta_{\CO}}$, with $Y$ a pure number. 
Hence, the series (\ref{2.6}) relative to its one--point function at
finite temperature can be cast in the scaling form 
\begin{equation}
\frac{\langle\CO\rangle_R}{\langle \CO\rangle_0} = 
H(m R)\,\,\, ,
\label{scaling1} 
\end{equation}
where $H(m R)$ is a function of the scaling variable $m R$. 

In the low--temperature limit $R \rightarrow \infty$, the pseudo--energy 
goes as $\epsilon (\th) \sim m R \cosh\th$ and therefore the 
$N$-th term of the series entering $H(m R)$ vanishes asymptotically 
as $e^{-N m R}$, so that the leading correction is given by  
\begin{equation}
H(m R) \rightarrow 1 + \frac{\langle A|\CO|A\rangle}{\pi} K_0(m R) 
+ \cdots  
\end{equation}
where $K_0(x)$ is the usual Bessel function, 
and $|A\rangle = | \theta \rangle$. 

On the other hand, in the high--temperature limit $R \rightarrow 0$, 
the one--point correlators may become scaling invariant 
functions\footnote{This is not always the case, because there are 
theories, like the thermal Ising model, with logarithmic behavior or
models, like the Sinh-Gordon one, which present logarithmic
corrections which may spoil the pure power law behavior given in 
the text.}
i.e.
\EQ
\langle \CO\rangle_R
\rightarrow
\frac{1}{R^{\eta}}\,\,\, , 
\label{scaling} 
\end{equation}
with a power--law exponent $\eta$ ruled by the underlying Conformal Field
Theory. This is the case, for instance, of massive integrable models 
obtained as a deformation of a Conformal Field Theory by a strong relevant
field. Under this hypothesis, the limit $R\rightarrow 0$ can be controlled 
by means of a Conformal Perturbation Theory on the cylinder.
Let  
\begin{equation}
{\cal A} = {\cal A}_{CFT} + \lambda \int \Phi(x) d^2x \,\,\, ,
\end{equation}
be the action of the off--critical model, where $\Phi(x)$ is the 
relevant operator which gives rise to the massive integrable theory, 
with $\lambda \sim m^{2 - 2\Delta_{\Phi}}$. The general structure of 
the perturbation series for the one--point functions is then given by 
\begin{equation}
\langle \CO \rangle_R =
\sum_{n=0}^n d_n \,\,\, ,
\label{series}
\end{equation}
where 
\EQ
d_n = \frac{(-\lambda)^n}{n!} \int_{cyl} \langle \xi | \CO(0) 
\Phi(X_1) \cdots \Phi(X_n) |\xi \rangle_{conn} \,d^2X_1 \cdots 
d^2X_n \,\,\, ,
\label{perturbative}
\EN 
and $X_i$ are points on the cylinder and the connected correlation 
functions are calculated in the unperturbed CFT. The field $\xi$ entering 
eq.\,(\ref{perturbative}) is relative to the conformal operator of lowest 
anomalous dimension in the theory which plays the role of the vacuum 
state on the cylinder (for a unitary model, this field is the identity 
operator). Notice that the first term of the series is given  by 
\EQ
\langle \xi | \CO | \xi\rangle = \left(\frac{2 \pi}{R}\right)^{2\Delta_{\CO}}
{\cal C}_{\xi\CO\xi}
\,\,\, ,
\label{lowest}
\EN 
and
this will be different from zero only if the conformal structure constant 
${\cal C}_{\xi\CO\xi}$ does not vanish. In this case, the exponent 
$\eta$ in (\ref{scaling}) will coincide with the anomalous dimension 
of the field $\CO$ itself, $\eta = 2\Delta_{\CO}$, otherwise the exponent
$\eta$ will be given by $\eta = 2\Delta_{\CO} - q(2-2\Delta_{\Phi})$, 
where the integer $q$ is the first non--vanishing term in the series 
(\ref{series}). 

The above considerations open up the possibility to extract conformal data 
out of the high--temperature limit of the one--point function. This is 
easily checked by the analysis of some simple models which have 
only one massive particle in the spectrum.  

\subsubsection{Yang--Lee model} 

The $S$--matrix of the perturbed Yang--Lee model was determined in 
\cite{CarMus} to be 
\EQ
S(\th)\,=\,
\frac{\tanh\frac{1}{2}(\th + i\frac{2\pi}{3})}
{\tanh\frac{1}{2}(\th - i\frac{2\pi}{3})} \label{smatrixYL}\,\,\,.
\EN
Therefore for $\varphi(\th)$ we have 
\EQ
\varphi(\th) = -\frac{\sqrt{3}\cosh\th}{\cosh^2\th -
\frac{1}{4}}\,\,\,.
\EN 
The only off--critical primary operator of the theory coincides 
with the trace of the stress--energy tensor. In the conformal 
limit it goes to the field with the lowest anomalous dimension $2\Delta =
-2/5$ which simultaneously plays the role of the vacuum of the cylinder. 
Hence, in this case we expect to find for the exponent $\eta =-2/5$. 
In order to extract this parameter, by taking the logarithm of 
both terms in (\ref{scaling})
\EQ
-\eta \log R = \lim_{R\rightarrow 0} \langle \CO\rangle_R \,\,\, 
\label{leading}
\EN 
$\eta$ is easily identified by isolating the terms proportional 
to $-\log R$ in the series of the right hand side. Observe that, 
factorizing the term $(T_\mu^\mu)_0$ in eq.\,(\ref{seriestrace}), 
the logarithm of the series (\ref{seriestrace}) can be expressed in terms 
of the usual cluster expansion
\begin{eqnarray}
\log \langle T_\mu^\mu\rangle_R - \log (T_\mu^\mu)_0 & = & 
\frac{2\pi m^2}{(T_\mu^\mu)_0} \left[
\int \frac{d\th}{2\pi} f(\th) e^{-\epsilon(\th)} 
+ \int\int \frac{d\th_1}{2\pi}\frac{\th_2}{2\pi} f(\th_1) f(\th_2) 
e^{-\epsilon(\th_1) -\epsilon(\th_2)} \times \,
\right. \nonumber\\
&& \left. \times 
\left[\varphi(\th_1-\th_2) \cosh(\th_1-\th_2) - \frac{1}{2} 
\left(\frac{2\pi m^2}{(T_\mu^\mu)_0}\right)\right] + \cdots \right]
\label{cluster1}
\end{eqnarray} 
where, for this model 
\EQ
\frac{2\pi m^2}{(T_\mu^\mu)_0} = - 2 \sqrt{3} \,\,\,.
\label{ratio1}
\EN 
Following the TBA analysis \cite{Alosh}, it is known that in the 
limit $R \rightarrow 0$, the pseudo--energy $\vep(\th)$ 
tends to a plateau of value $\epsilon_0 = \log\left[
\frac{\sqrt{5} +1}{2}\right]$ in a region $|\th| < 2 \log(2/R)$. 
Therefore the first estimate of $\eta$ is obtained by the first 
term of the right hand side of (\ref{cluster1}) 
\begin{eqnarray}
{\cal K}_1 & = & 
\frac{2\pi m^2}{(T_\mu^\mu)_0} 
\int \frac{d\th}{2\pi} 
f(\th) 
e^{-\epsilon(\th)} \,\simeq\,
\frac{2\pi m^2}{(T_\mu^\mu)_0} 
\int_{-\log\left(\frac{2}{r}\right)}^ 
{\log\left(\frac{2}{r}\right)} f(\th) 
e^{-\epsilon(\th)} = \\
&& \,\,\, \frac{2\pi m^2}{(T_\mu^\mu)_0} \frac{1}{1+e^{\epsilon_0}} 
\frac{1}{\pi} \log\left(\frac{2}{R}\right) \,\,\, ,
\nonumber
\end{eqnarray}
i.e. 
\EQ
\eta \simeq - \frac{1}{\pi} \frac{4 \sqrt{3}}{3 + \sqrt{5}} = 
-0.421178...
\label{1app}
\EN 
which is  within $5\%$ to the actual value $\eta = -0.4$. 
The next correction is obtained by isolating the term proportional 
to $-\log R$ in the next term of the cluster expansion 
(\ref{cluster1}), which is equivalent to computing the quantity 
\EQ
\frac{3}{\pi^2} \left(\frac{1}{3 + \sqrt{5}}\right)^2 
\int_{-\infty}^{+\infty} d\th \frac{1}{\cosh^2\th -\frac{1}{4}} 
=0.0268126...
\EN 
so that the estimate of $\eta$ improves to the values 
\EQ
\eta \simeq - 0.394365...
\EN 
which differs from the actual value by $1\%$.

The main result of these computations can be summarized by saying 
that the exponent $\eta$ will be expressed by a series (of alternating 
signs, for this model) whose terms involve, in addition to the 
ratio (\ref{ratio1}), inverse powers of $\pi$ and the plateau 
value $\epsilon_0$.  

\subsubsection{The non--unitary model ${\cal M}_{3,5}$}

Another integrable model with only one massive excitation to which we can
easily apply the above consideration is the off--critical
``thermal'' deformation of the minimal model ${\cal M}_{3,5}$, which 
has been previously studied in \cite{Smirnov35,DM,CM}. The 
$S$--matrix of the model is given by $S(\th) = 
-i \tanh\frac{1}{2}\left(\th - i \frac{\pi}{2}\right)$ and therefore 
$\varphi(\th) =1/\cosh\th$. In the conformal limit, the trace 
of the stress--energy tensor tends to the conformal field $\CO$ of 
anomalous dimension $2\Delta = 2/5$ and this field has a non--vanishing
structure constant with the field $\xi$ of anomalous dimension 
$2\Delta_{\xi} = -1/10$ which plays the role of vacuum state of the 
cylinder. Hence we expect that the exponent $\eta$ of this theory 
should be equal to $2/5$. By using the same arguments as above, 
with the additional data  
\EQ
\frac{2\pi m^2}{(T_\mu^\mu)_0} =  2 \,\,\,\,\, , ~~~~~~~~~ 
\epsilon_0 = -\log\left[\frac{\sqrt{5}+1}{2}\right] \,\,\, ,
\label{ratio2}
\EN 
for the first estimate of the exponent $\eta$ we have 
\EQ
\eta \sim \frac{2}{\pi} \frac{1+\sqrt{5}}{3+\sqrt{5}} = 0.393453... 
\EN 
which is within $1\%$ of the  actual value $\eta = 0.4$. 
In this model, the fact that the first term of the cluster series
gives a better approximation to $\eta$ seems to be related to 
the vanishing of the next leading correction in the cluster 
expansion. 

\section{Two-Point Functions}

Having understood the main features of the one-point functions, 
the same reasoning applies
to the higher point functions.  In this section we focus on the
two-point functions.  We begin with the double summation over
states
\begin{equation}
\label{3.1}
\langle \CO(x,t) \CO (0) \rangle_R = \inv{Z} \sum_{\psi, \psi'} 
e^{-E_\psi R} 
\langle \psi | \CO (x,t) |\psi' \rangle \langle \psi' | \CO (0) 
| \psi \rangle \,\,\, .
\end{equation}
As explained in \cite{LLSS}, for a free theory the terms in 
(\ref{2.4}) with $\CA_1 \neq \CA$ give rise to three kinds of
contributions:

\noindent
(i).  Terms that diverge in infinite volume involving 
$[\delta(\th - \th')]^2$ which can be arranged into an overall factor of $Z$. 

\noindent
(ii). Terms that give rise to factors of the one-point function 
$( \langle \CO \rangle_R)^2$.  

\noindent 
(iii).  Terms that modify the integration over rapidity for
{\it both} the $\psi$ and $\psi'$ states by the same filling fraction $f$. 

\noindent
The remaining contributions are only expressed in terms of connected 
form-factors for which crossing symmetry is valid.  As for the
one-point function, the only difference between the free and
interacting theory is that $e,k$ should be replaced by the TBA
dressed quantities $\tilde{e}, \tilde{k}$.  A further justification 
of this will be given below.

\def\sigt{{\tilde{\sigma}}}

Introduce an index $\sigma = 1$ for the $\psi'$ particles and
$\sigma = -1$ for the $\psi$ particles.  For the $\psi'$ particles
the integrations $\int d\th$ are accompanied by $f(\th)$ defined
in (\ref{2.7}), whereas for the $\psi$ particles one has 
$f(\th) e^{-\vep (\th)}$ due to the $e^{-E_\psi R}$ in (\ref{3.1}).  
These factors can be expressed as $f_\pm (\th)$ as defined in 
(\ref{fillsig}). 
By using the crossing symmetry, the connected form factors appearing 
in (\ref{3.1}) are 
$\langle \psi' | \CO(0) | \psi \rangle_{\rm conn}  = 
\langle \psi + i\pi , \psi' | \CO (0) | 0 \rangle$, where
$\psi + i\pi$ denotes all rapidities shifted by $i\pi$.  
We can describe this by defining 
\begin{equation}
\label{3.3}
\langle 0 | \CO | \th_1 \cdots \th_N \rangle_{\sigma_1 \cdots \sigma_N} 
= \langle 0 | \CO | \th_1 - i\pi\sigt_1 , 
\cdots \th_N - i\pi \sigt_N \rangle \,\,\, ,
\end{equation}
where $\sigt = (\sigma -1 )/2  \in \{ 0, 1 \}$.  Finally,
\begin{eqnarray}
\label{3.4}
\langle \CO (x,t) \CO (0) \rangle_R &=& (\langle \CO \rangle_R)^2 
\sum_{N=0}^\infty \inv{N!} \sum_{\sigma_i = \pm 1} 
\int \frac{d\th_1}{2\pi} \cdots \frac{d\th_N}{2\pi}
\[ \prod_{j=1}^N f_{\sigma_j} (\th_j ) \exp \( -\sigma_j (t \tilde{e}_j 
+ i x \tilde{k}_j )  \)  \] 
\\ \nonumber
&~&
~~~~~~~\times 
\left\vert \langle 0 | \CO (0) | \th_1 \cdots \th_N \rangle_{\sigma_1 
\cdots \sigma_N } \right\vert^2 \,\,\, ,
\end{eqnarray}
where $\tilde{e}_j , \tilde{k}_j = \tilde{e} (\th_j ) , \tilde{k} (\th_j )$. 

\def\thl{{\overleftarrow{\theta}}}
\def\thr{{\overrightarrow{\theta}}}

A non-trivial check of the above expression is the Kubo-Martin-Schwinger
formula \cite{KMS}, which for a bosonic operator $\CO$ reads 
\begin{equation}
\label{3.5}
\langle \CO (x, t+R) \CO (0) \rangle_R = \langle \CO (0) \CO (x,t) 
\rangle_R \,\,\, . 
\end{equation} 
Shifting $t$ by $R$ in (\ref{3.4}) gives the factor $e^{-\sigma \vep}$. 
Noting that $f_\sigma e^{-\sigma \vep} = f_{-\sigma} $, we simply
relabel the particles $\sigma \to - \sigma$ in the sum.  
Since $\langle \CO (0) \CO (x,t) \rangle = \langle \CO (-x, -t) \CO (0) 
\rangle$, then (\ref{3.5}) holds as long as 
\begin{equation}
\label{3.6} 
\left\vert \langle 0 | \CO (0) | \th_1 \cdots \th_N \rangle_{\sigma_1
\cdots \sigma_N } \right\vert^2
=
\left\vert \langle 0 | \CO (0) | \th_1 \cdots \th_N \rangle_{-\sigma_1
\cdots -\sigma_N } \right\vert^2 \,\,\, .
\end{equation}

The latter identity can be proven as follows.  
Let $| \thr - i\pi \sigt \rangle$ denote 
$| \th_1 - i\pi \sigt_1 , \cdots, \th_N - i\pi \sigt_N \rangle$ 
and $\langle \thl + i\pi \sigt |$ denote 
$\langle \th_N + i\pi \sigt_N , \cdots , \th_1 + i \pi \sigt_1 |$.   
Then,
\begin{equation}
\label{3.7}
\left\vert \langle 0 | \CO (0) | \th_1 \cdots \th_N \rangle_{\sigma_1
\cdots \sigma_N } \right\vert^2
= \langle 0 | \CO | \thr - i\pi \sigt \rangle 
\langle \thl + i\pi \sigt | \CO | 0 \rangle \,\,\, . 
\end{equation}
Under the transformation $\sigma \to - \sigma$, one has
$\sigt \to - \sigt -1 $.  For local operators, the form factor is
invariant under a shift of all rapidities by the same constant, 
thus
\begin{eqnarray}
\label{3.8}
\left\vert \langle 0 | \CO (0) | \th_1 \cdots \th_N \rangle_{-\sigma_1
\cdots -\sigma_N } \right\vert^2
&=& 
\langle 0 | \CO | \thr + i\pi \sigt \rangle 
\langle \thl - i\pi \sigt | \CO | 0 \rangle 
\\ \nonumber
&=&
\langle 0 | \CO | \thl - i\pi \sigt \rangle 
\langle \thr + i\pi \sigt | \CO | 0 \rangle \,\,\, , 
\end{eqnarray}
where we have used crossing symmetry in the second line. 
The orders of rapidities in (\ref{3.8}) can be brought to the order
in (\ref{3.7}) using 
$ \langle 0 | \CO | \cdots \th_i ,\th_j , \cdots \rangle 
= S(\th_i - \th_j ) \langle 0 | \CO | \cdots \th_j ,\th_i , \cdots \rangle
$.  In doing this one always encounters the pair of factors 
$S(\th - i\pi ) S(-\th - i\pi)$.  Using crossing $S(\th) = S(i\pi - \th)$
and unitarity $S(\th) S(-\th) = 1$, the latter factor equals $1$.  
Thus, (\ref{3.6}) holds.

\section{Conclusions}

We have proposed explicit integral representations for the one and
two-point correlation functions at finite temperature involving
form factors and the TBA pseudo-energy-momentum.  Several checks
were performed, in particular we showed that for the trace
of the energy momentum tensor, our expression for the
one-point function  coincides precisely with the TBA result.  
We also checked that the Kubo-Martin-Schwinger relation holds 
for the two-point function.

We close by listing some possible applications of our formalism 
that deserve further investigation. By extending our results 
to boundary field theories at finite temperature, as was done 
for the free case in \cite{LLSS}, it should be possible to 
study the finite temperature properties of conductivities in 
quantum wires with impurities. Another application is to 
the study of finite temperature crossovers in the vicinity 
of a zero temperature quantum phase transition (for a review 
see \cite{Subir}). Finally there is the issue of spin-diffusion 
in quantum spin chains, which would require the study of the 
infinite temperature limit of the continuum version of a 
spin-spin two-point correlation function\cite{mccoy}.

\vspace{1cm}
{\em Acknowledgments}. We would like to thank F. Lesage for discussions.
We are also grateful to P. Simon for his numerical support in producing 
Figure 1.

\newpage

{\large{\bf{Appendix A. TBA Dressed Energy and Momentum}}} 

\vspace{3mm}
Let $(n_j , \th_j )$ and $(n'_j , \th'_j )$ be two Bethe states 
both satisfying (\ref{2.9}), where $n'_j = n_j$ except for $j=\alpha$. 
Subtracting the two equations one obtains 
\begin{equation}
\label{A1} 
mL \( \sh \th'_j - \sh \th_j \) = \sum_i \sigma(\th_j - \th_i ) 
- \sigma (\th'_j - \th'_i ) \,\,\, .
\end{equation}
$(j\neq \alpha)$.  Since $\th'_j \approx \th_j $ we can introduce
a function $\chi (\th )$ and write 
\begin{equation}
\label{A2}
L \( \sh \th'_j - \sh \th_j \) \approx \chi (\th_j ) \ch \th_j \,\,\, . 
\end{equation}
In this thermodynamic limit, (\ref{A1}) can be written as 
\begin{equation}
\label{A3}
2\pi ( 1- \hat{K} ) (\rho \chi ) = \sigma (\th - \th_\alpha ) 
- \sigma (\th - \th'_\alpha) \,\,\, ,
\end{equation}
where $\rho$ is the previously introduced density of levels
and $\hat{K}$ is defined in (\ref{2.21}). 
One also has 
\begin{equation}
\label{A4}
\sigma (\th - \th_\alpha ) 
- \sigma (\th - \th'_\alpha) 
= \int_{\th_\alpha}^{\th'_\alpha} d\th' 
\hat{K} (\th , \th' ) ( 1 + e^{\vep (\th')} ) \,\,\, . 
\end{equation}
Using the resolvent operator, as defined in (\ref{2.23b}), one has 
\begin{equation}
\label{A5} 
\rho \chi (\th ) = \int_{\th_\alpha}^{\th'_\alpha} 
\frac{d\th'}{2\pi}  \hat{L} (\th , \th') ( 1 + e^{\vep (\th')} )
\,\,\, .
\end{equation}
Now consider the difference in energy $\Delta E$ between the two
Bethe states:
\begin{equation}
\label{A6}
\Delta E = m \,  \ch \th'_\alpha - m \, \ch \th_\alpha 
+ m \int d\th ~ \sh \th ~ \frac{\chi (\th) \rho (\th )}{ 1 + e^{\vep (\th )} } 
\,\,\, .
\end{equation}
Substituting (\ref{A5}), and using the property 
$ \hat{L} (\th , \th') ( 1 + e^{\vep (\th')} ) 
= ( 1 + e^{\vep (\th )} ) \hat{L} (\th' , \th ) $, 
along with (\ref{2.23c})  one finds 
\begin{equation}
\label{A7} 
\Delta  E = \inv{R} \( \vep (\th'_\alpha ) - \vep (\th_\alpha ) \) 
\,\,\, .
\end{equation}

The dressed momentum proceeds similarly: 
\begin{eqnarray}
\label{A8} 
\Delta P &=& \sum_j m \, \sh \th'_j  - m \, \sh \th_j = 
\\ \nonumber
&=& 
m\,  \sh \th'_\alpha - m \,  \sh \th_\alpha 
+  
m \int d\th ~ \ch \th ~  \frac{\chi (\th) \rho (\th )}{ 1 + e^{\vep (\th )} } 
\,\,\,.
\end{eqnarray}
Again substituting (\ref{A5}) and using (\ref{2.23c}) one finds 
\begin{equation}
\label{A9} 
\Delta P = \tilde{k} (\th'_\alpha ) - \tilde{k} (\th_\alpha) 
\end{equation}
where $\tilde{k}$ is defined in (\ref{2.17}).

\newpage
\begin{center}
\begin{minipage}[b]{.65\linewidth}
\centerline{\epsfxsize=12.cm\epsfbox{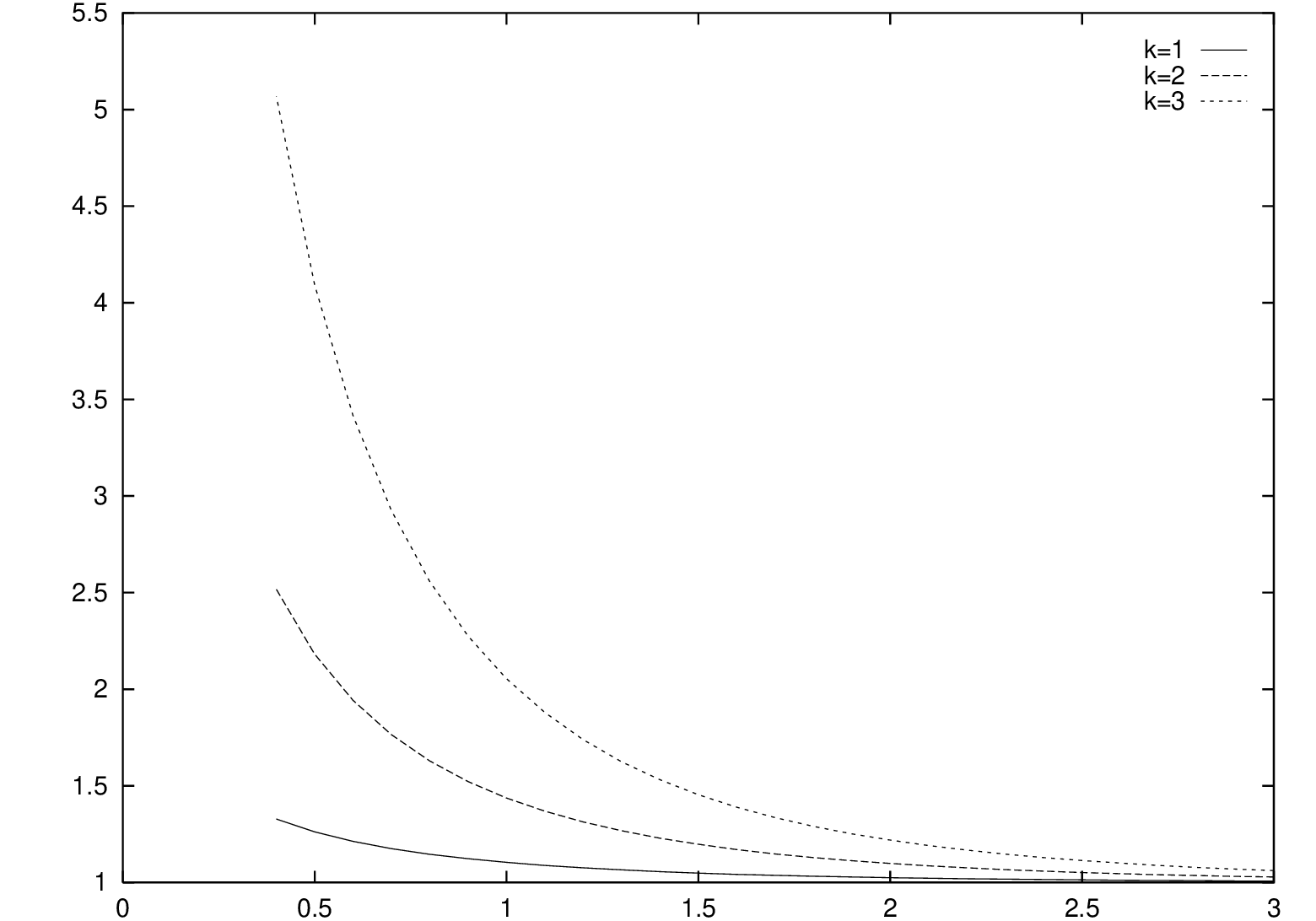}} 
\end{minipage}
\end{center}

\vspace{5mm}
\begin{center}
Figure 1a. One--point correlators of $e^{k g \Phi}$ as functions 
of $m R$ for $\alpha=1/20$.
\end{center}

\vspace{15mm}

\begin{center}
\begin{minipage}[b]{.65\linewidth}
\centerline{\epsfxsize=12.cm\epsfbox{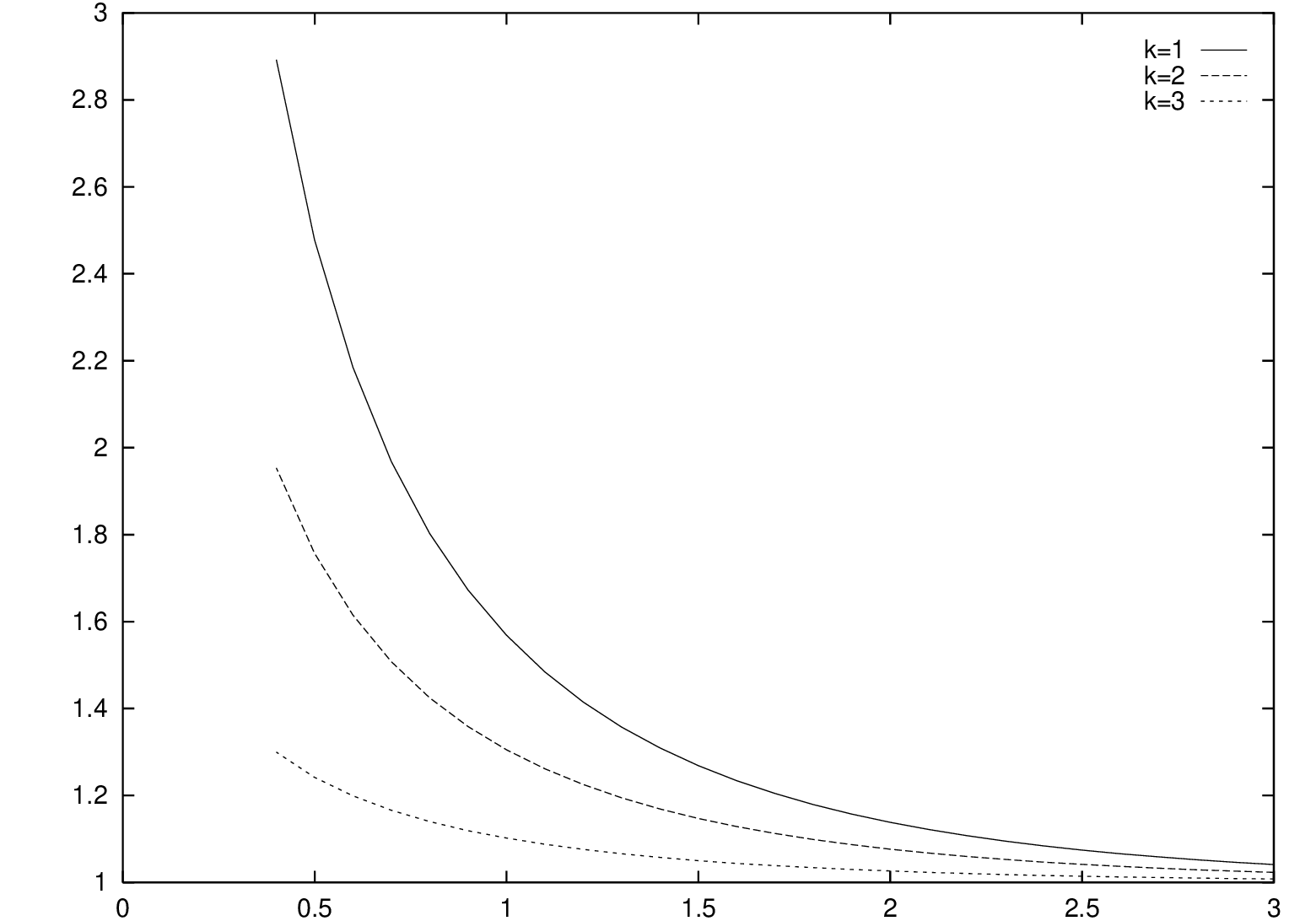}} 
\end{minipage}
\end{center}

\vspace{5mm}
\begin{center}
Figure 1b. One--point correlators of $e^{k g \Phi}$ as functions 
of $m R$ for $\alpha=1/\sqrt{7}$.
\end{center}


\end{document}